\documentclass[a4paper,fleqn]{cas-sc}

\usepackage{pdfpages}
\usepackage[
  separate-uncertainty = true,
  multi-part-units = repeat
]{siunitx}
\usepackage{longtable}
\usepackage{setspace}
\usepackage{array}
\usepackage{ragged2e}

\usepackage[numbers]{natbib}
\def\tsc#1{\csdef{#1}{\textsc{\lowercase{#1}}\xspace}}
\tsc{WGM}
\tsc{QE}
\tsc{EP}
\tsc{PMS}
\tsc{BEC}
\tsc{DE}

\begin{document}
\let\WriteBookmarks\relax
\def\floatpagepagefraction{1}
\def\textpagefraction{.001}
\shorttitle{A review on plasmonic and metamaterial based biosensing platforms for virus detection}
\shortauthors{M M Hassan et~al.}


\title [mode = title]{A review on plasmonic and metamaterial based biosensing platforms for virus detection}                      

\author[1,2]{Mohammad Muntasir Hassan}[orcid=0000-0001-7235-0710]
\ead{hmuntasir@iict.buet.ac.bd}

\address[1]{Department of Electrical and Electronic Engineering, Bangladesh University of Engineering and Technology, Dhaka.}

\author[1,3]{Farhan Sadik Sium}[orcid=0000-0002-9784-8256]
\ead{sium1093@gmail.com}

\author[1,4]{Fariba Islam}[orcid=0000-0001-6125-4993]
\ead{faribaislam3011@gmail.com}

\address[2]{Institute of Information and Communication Technology, Bangladesh University of Engineering and Technology, Dhaka.}
\address[3]{Department of Electrical and Electronic Engineering, Daffodil International University, Dhaka.}
\address[4]{Department of Computer Science and Engineering, BRAC University, Dhaka.}
\author%
[1]{Sajid Muhaimin Choudhury}
\cormark[1]
\ead{sajid@eee.buet.ac.bd}

\cortext[cor1]{Corresponding author}

\begin{abstract}
\doublespacing
Due to changes in our climate and constant loss of habitat for animals, new pathogens for humans are constantly erupting. SARS-CoV-2 virus, become so infectious and deadly that they put new challenge to the whole technological advancement of healthcare. Within this very decade, several other deadly virus outbreaks were witnessed by humans such as Zika virus, Ebola virus, MERS-coronavirus etc and there might be even more infectious and deadlier diseases in the horizon. Though conventional techniques have succeeded in detecting these viruses to some extent, these techniques are time-consuming, costly, and require trained human-resources. Plasmonic metamaterial based biosensors might pave the way to  low-cost rapid virus detection. So this review discusses in details, the latest development in plasmonics and metamaterial based biosensors for virus, viral particles and antigen detection and the future direction of research in this field.  

\end{abstract}



\maketitle
\doublespacing

\section{Introduction}  
In late 2019, an unprecedented case of pneumonia was diagnosed in China which later was proved to be caused by a novel severe acute respiratory syndrome -coronavirus (SARS-CoV-2) or novel coronavirus\cite{huang2020clinical, lu2020genomic}. This novel coronavirus disease (COVID-19) spread throughout the world in a very short time and was declared a pandemic by the World Health Organization (WHO)\cite{world2020director}. It is to be noted that this is the third large-scale outbreak of Coronavirus associated disease within less than a decade after Severe Acute Respiratory Syndrome (SARS) in 2003\cite{zhong2003epidemiology} and Middle East Respiratory Syndrome (MERS) in 2012\cite{alhamlan2017case}. Apart from novel Coronavirus, a number of other virus related diseases also cause wreak havoc to the global economy and it is needless to say that a rapid, reliable and accurate detection of viruses can contribute greatly to control the spread of the disease and prevent future pandemics like COVID-19. 

Currently, several common methods are used for detecting infectious viruses. Serological testing \cite{bastos2020diagnostic}, immunofluorescence, nucleic acid amplification test (NAAT) are the most common genre of diagnosis \cite{vemula2016current}. Hemagglutination inhibition assay (HI) and enzyme linked immunosorbent assay (ELISA) are the major serological testing \cite{souf2016recent}.  However, they have some major drawbacks which have hindered their ubiquitous usage for virus detection. For instance, the preparation of antibody for ELISA is a costly technique and requires expert manpower\cite{sakamoto2018enzyme}. On the other hand, HI has low specificity under certain levels of agglutination and the sample may contain non-specific hemagglutinating factors \cite{soares1999detection}. Nucleic acid amplification based reverse transcription polymerase chain reaction (RT PCR) test is another technique to detect virus and to date, it is the most widely used technique to detect novel Coronavirus\cite{corman2020detection}.  RT–PCR technique is a highly sensitive, specific and reliable diagnostic method. But this test typically takes longer than other detection methods and requires expert manpower and hence is expensive. Nucleic acid sequence‐based amplification (NASBA), loop‐mediated isothermal amplification (LAMP), RT–PCR and q-PCR are in the genre of nucleic acid amplification test (NAAT). Most of these testing schemes are costly and require a lot of time and highly trained manpower. Other methods to detect viruses such as CRISPR \cite{qin2019rapid} and culture methods have apparently failed to be a rapid and reliable method for satisfactory diagnosis of different viruses as they are not widely used yet. In such circumstances, real-time and label-free biosensors have recently emerged as auspicious diagnostic tools for different infectious diseases. These sensors overcome the need for fluorescence or radioactive tagging for virus detection, thus enabling compact, robust, cost-effective point-of-care diagnostics. Different bio-sensing platforms based on optical, electrical \cite{luo2013electrical}, and mechanical \cite{savran2004micromechanical} has shown promising applications ranging from laboratory investigation to clinical diagnostics and drug development to combating emerging infectious diseases. Among these different genres of biosensors, optical detection platforms have gained considerable interest in recent years. Optical biosensors allow remote diagnosis scheme of the bio-molecular binding signal from the sensing volume without any physical connection between the excitation source and the detection media. Unlike mechanical and electrical sensors, these optical sensors are also compatible with physiological solutions and are not sensitive to the changes in the ionic strengths of the solutions. Among different optical biosensors, plasmonic and metamaterial based plasmonic biosensors are highly potential in this regard due to their exotic properties like miniaturized sensor chip\cite{yesilkoy2018phase}, real-time sensing \cite{guner2017smartphone}, label-free sensing mechanism \cite{maalouf2007label}. 

The aim of this comprehensive review is to present the advances in plasmonic and metamaterial based plasmonic biosensors for virus or viral particles detection and highlight the scopes of future work in this field. There have been some recent review papers on plasmonic biosensors for virus detection \cite{mauriz2020recent,shrivastav2021comprehensive}, different methods of Coronavirus detection \cite{samson2020biosensors,yuce2020covid,ji2020detection,antiochia2020developments} and recent progress in nanophotonic biosensors to combat the COVID-19 pandemic \cite{soler2020nanophotonic}. However, metamaterial based plasmonic biosensors haven't been covered before and in this review, metamaterial based virus detection methods are discussed in detail which is unprecedented and different plasmonic biosensors are classified in five broad fields based on the detection technique and structure of the biosensors. Due to the ongoing COVID-19 pandemic, emphasis is given to the family of Coronavirus detection techniques and the performance of a number of biosensors for various virus detection are also compared and summarized.  Moreover, we have discussed the future trends in plasmonic and metamaterial based biosensing such as surface plasmon resonance imaging (SPRi), incorporating quantum properties of materials in biosensing, novel materials based biosensors, artificial intelligence, and machine learning application in biosensing. As a non-destructive virus sensing platform, the potential application of plasmonic and metamaterial based biosensors for rapid, multiplexed, point-of-care detection of virus is also highlighted.


\begin{figure}  

  \centering  

        \includegraphics[trim={0cm 2cm 0cm 2cm},clip,scale=.54]{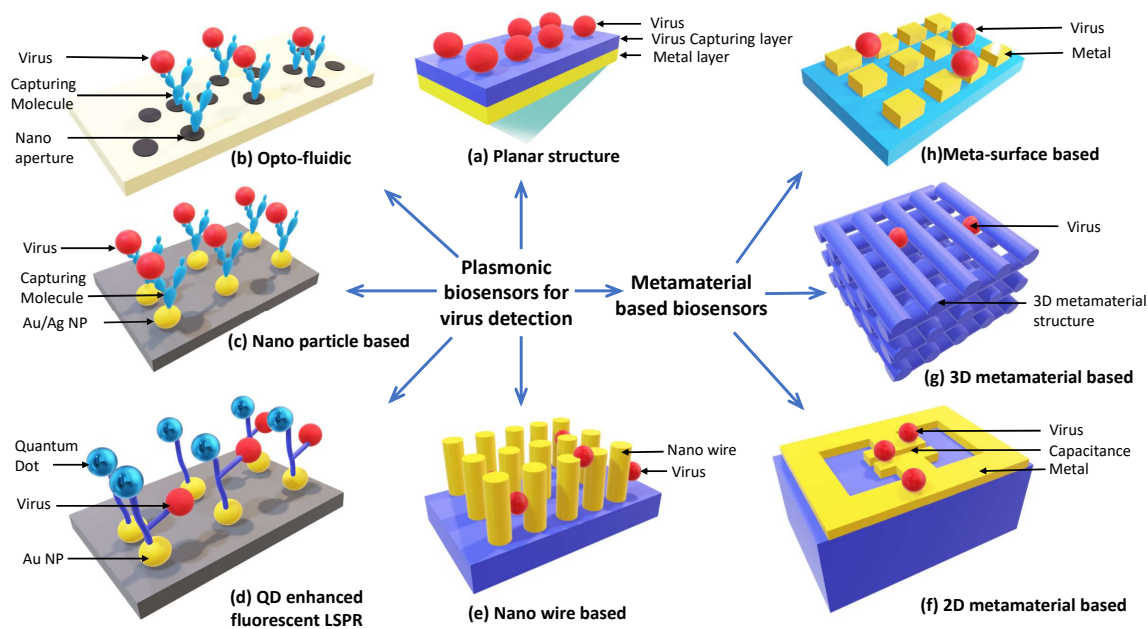}  

    \caption{Schematic of different plasmonic and metamaterial based virus-sensing structures. (a) Planar structure: Surface plasmon is generated in between dielectric and metal. (b) Opto fluidic structure: nano aperture holding antibody increasing binding potential for flowing virus antigen. (c) nanoparticle: localized surface plasmon around NPs enhances the sensitivity. (d) Quantum dots attachment with NPs: Binding QD with NP enables enhanced fluorescent LSPR sensing. (e) Nano wire: Plasmons are generated around nano wires increasing the sensitivity. (f) 2D metamaterial: virus attachment in metamaterial changes capacitance which changes the resonant frequency. (g) 3D metamaterial: 3D shaped metamaterial can mend magnetic field of light more efficiently which has the potential to materialize ultra-sensitive biosensors. (h) Metasurface: unusual patterns of metasurface performs as an efficient virus sensing platform.}

\end{figure}

\section{Evaluation of plasmonic biosensors}  

To evaluate the performance of biosensors several figures of merit are widely used. Among them, detection limits or limit of detection (LOD), sensitivity, selectivity or specificity are the most popular. Selectivity or specificity (S) is defined as the ability of a sensor to detect a particular virus from a sample containing admixtures of similar or other materials.  Sensitivity and detection limits are two significant metrics that can be used to compare biosensors of different platforms. Sensitivity in the case of virus detection expresses how a sensor interacts in the presence of the virus. For virus detection generally sensitivity of plasmonic biosensor is defined as-

\begin{equation}
S = \frac{\Delta \lambda}{\Delta C}
\end{equation}  

Here $\Delta \lambda$ is the change in peak reflection/transmission wavelength and $\Delta C$ is the change in virus concentration.  

Another important figure of merit limit of detection (LOD) or detection limit (DL) is defined as the minimum virus concentration that can be detected by the sensor. In other words, LOD is the minimum number of viruses necessary to cause a detectable change in the output signal of the sensor. For the determination of LOD a formula commonly used is- \cite{shrivastava2011methods,suthanthiraraj2019localized}

    \begin{equation}
    LOD = \frac{3\sigma}{S}
    \end{equation}

Here $\sigma$ is the standard deviation of the control without virus which is basically the system noise floor and S is the slope of the linear fit for wavelength shift versus virus concentration plot which is the sensitivity of the sensor. LOD for virus detection can be specified in different units. But commonly used units are: (1) ng/mL (2) copies of virus/mL (3) PFU/mL  (4) pg/$mm^2$ (5) EID/mL

\section{From plasmonic to metamaterial based biosensor}
\subsection{Plasmonic excitation in biosensing}
Metal-dielectric contact has been one of the primary methods of generating plasmonic excitation. Generally, it is a guided mode that propagates along metal/dielectric interfaces. Plasmonic excitations are characterized into two segments, namely surface plasmon polaritons (SPP) or surface plasmon resonance (SPR) and localized surface plasmon resonance (LSPR). SP that propagates at the flat interface between a conductor and a dielectric is a two-dimensional electromagnetic wave. It is the collective resonant oscillation of conduction electrons and incoming photons at the interface between the metal and the dielectric as shown in Figure 1(a). On the other hand, LSPR is generated by a light wave trapped within conductive nanoparticles (NPs) smaller than the wavelength of light as depicted in Figure 1(c). The size of the NPs is typically in the region of Rayleigh scattering. When an external electric field is applied to metallic NPs, the conduction electrons encounter combined harmonic oscillations causing a strongly localized electromagnetic field with high intensity. Ever since the discovery of this unique characteristic, many exciting research on biosensing has been conducted using this exotic property and it has been used in virus detection as well. Viruses like HIV, coronavirus, influenza, dengue, adeno virus, zika virus, hepatitis, norovirus etc. have been reported to be successfully detected by employing a variety of plasmonic biosensors.  

To induce SPR in the boundary between metal and dielectric, the momentum of the incident photon must be matched with the momentum of the conduction band electrons. If the matching condition is met, light can be coupled in the interface between the metal and dielectric plane. For flat planar surfaces, this phase matching is fulfilled by the attenuated total reflection (ATR). This usually requires a media of higher refractive index (RI). The matching condition can be interpreted from the dispersion relation given below \cite{raether1988surface}: 
\begin{equation} 
K_{spp}= \frac{2\pi}{\lambda_o} \sqrt{\frac{\varepsilon_m \varepsilon_d }{\varepsilon_m+ \varepsilon_d}} =\frac{2\pi}{\lambda_o} n_p sin\Theta _i
\end{equation} 

where  $n_p$ is the refractive index of the coupling prism, $\Theta_i$ is the incident angle of light, $\varepsilon_m$ is the dielectric constant of metal, $\varepsilon_d$ is the dielectric constant or dielectric permittivity, $K_{spp}$ is the wave vector of surface plasmon polariton. 

As the refractive index of the analyte media changes, $\varepsilon_d$  also changes, eventually altering the wave vector k. When $\varepsilon_m$  and $\varepsilon_d$ are equal and opposite of each other, the wave vector is maximum which results in resonance. Here $\varepsilon_m$ depends on the wavelength of incident light and $\varepsilon_d$ depends on the refractive index of the dielectric environment. Diverse configurations are used to generate SPR or LSPR for bio-sensing. In this review, these plasmonic biosensors are broadly classified into five different groups based on their structure and sensing principle namely planar structure, optofluidic structure, nanoparticle based structure, quantum dot based structure and nano rod-based structure as shown in Figure 1(a)-(e). 
 
\subsection{Emergence of metamaterials in biosensing} 
In recent years, to increase the sensitivity of plasmonic biosensors metamaterial based plasmonic biosensors have been employed. Advantages of using metamaterial based sensors are that a variety of geometric structures and different sensing principles can be utilized which were not feasible with conventional plasmonic biosensors.

In 1968, Russian physicist Victor Veselago first came up with the theoretical concept of left-handed materials \cite {veselago1968electrodynamics} which shows unusual refraction of light. Then in 1999, Pendry et al. theoretically showed that micro-structures built from nonmagnetic conducting sheets which are much smaller than the wavelength of radiation exhibit an effective magnetic permeability and these structures can be tuned to show varying magnetic permeability \cite{pendry1999magnetism} including imaginary component. Rodger Walser termed this type of substance as metamaterials in 1999. Smith with his colleagues experimentally demonstrated the first left-handed metamaterial at microwave frequency in the year 2000 \cite{smith2000composite}.  From then on metamaterials have been explored extensively for possible applications in optics \cite{pendry2000negative}, photonics \cite{fang2020mid}, energy harvesting \cite{yu2019broadband}, communication \cite{turpin2014reconfigurable}, sensing \cite{chen2012metamaterials,beruete2020terahertz}, biological imaging and spectroscopy \cite{zhou2019biological}. 
Primarily, electromagnetic metamaterials are utilized in biosensing applications and metamaterials based biosensors can be classified in different groups based on their structure- two dimensional, three dimensional and two dimensional meta-surface based biosensors. Among them, two dimensional meta-surface biosensors have been successfully employed in virus sensing.
Though usage of metamaterial in biosensing is still in its early stage, already detection of viruses like HIV\cite{ahmed2020tunable}, Zika virus \cite{ahmadivand2018extreme}, Avian Influenza Virus \cite{lee2017nano}, CPMV \cite{sreekanth2017hyperbolic}, PRD1 \cite{park2017sensing} using metamaterial based sensors have been reported and sensitivity has improved by approximately an order of magnitude \cite{hong2018enhanced}. Metamaterial based plasmonic biosensors have the potential to be a game changer in the field of label-free point of care virus detection.

\section{Sensing using plasmonic biosensors}

\begin{figure} 

  \centering 

        \includegraphics[scale=.54]{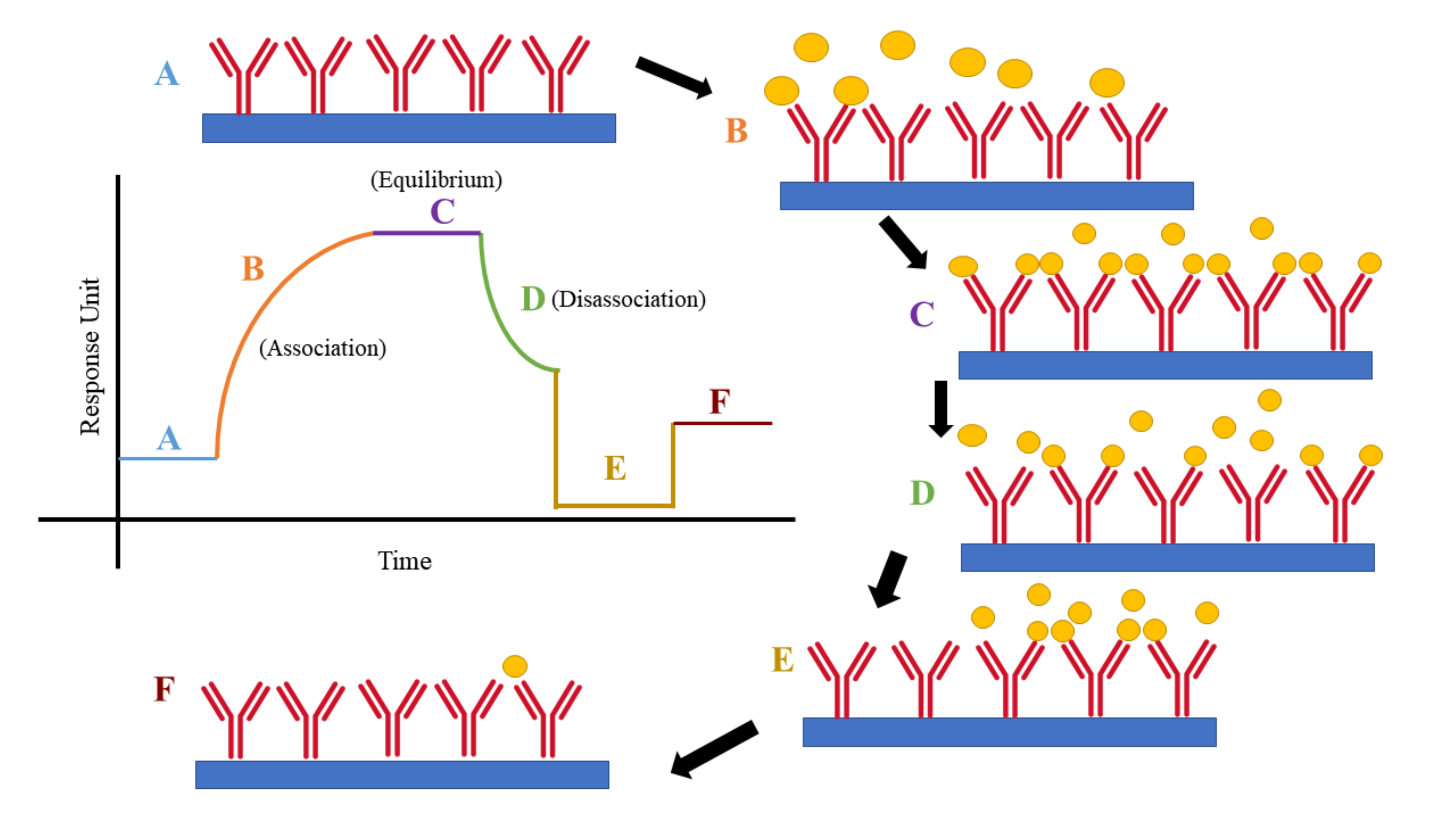} 

    \caption{Different stages of a bio-molecular reaction which can be observed by a planar SPR sensor. (a) Immobilized capturing molecules are attached to a sensor chip surface. (b) Association: pumped viral surface protein starts to bind with capturing molecule. (c) Equilibrium: almost all binding molecules captured the virus; (d) Disassociation: after certain time viral bodies starts to unbind with capturing probes. (f) With proper treatment the chip is ready for reuse.} 
\end{figure} 

Biosensor-based detection methods always utilize a specific bio-receptor surface to analyze either intact viruses or viral proteins. A common and widely explored bio-receptor is antibodies that originate in animal bodies against specific viral surface proteins or antigens. In addition to that, a number of artificial capturing molecules are developed in the laboratory to capture certain viruses such as laboratory made DNA (deoxyribonucleic acid) or RNA (ribonucleic acid) aptamers \cite{d2012artificial} and hairpin type mRNA \cite{liu2010quantitative}. The SPR system authorizes the characterization of the binding kinetics of these biomolecules in real-time. To analyze the reaction between biomolecules, generally, one interacting molecule is immobilized on the surface of the sensor chip, and its binding counterpart (sample analyte) is injected constantly into the solution through the flow cell, resulting in the analyte flowing over the capturing surface. As a result of the analyte reaction with the binding molecule, the analyte accumulates on the surface and increases the refractive index. The change in refractive index is measured in real-time, generating a plot of the response unit (RU) versus time. This entire process is depicted in Figure 2. The resulting responses obtained at different analyte concentrations are integrated to derive the rate constants (association, $K_a$; dissociation,$K_d$; and equilibrium, $K_e$). Thus, by using the SPR signal, the amount, and condition of analyte in the sample are diagnosed. SPR signal is usually measured in two ways for planar structures. Firstly, the change of incident angle with respect to the generation of SPR. Secondly, change of wavelength about SPR occurrence. SPR sensors are also categorized in this regard as incident angle modulated SPR sensor \cite{zhou2017angle} and wavelength modulated SPR sensor \cite{liu2005wavelength} respectively. In the following subsections, different plasmonic biosensors for virus detection are discussed in five broad categories.

\subsection{Planar structure}
SPR based planar structure biosensors are conventional plasmonic biosensors and recently they  are in the spotlight due to their ease of fabrication. They have been studied extensively for different virus detection. In 2020 Dengue Virus Type (DENV) 2E-proteins with high sensitivity and accuracy were successfully detected \cite{omar2020sensitive}. An SPR sensor based on self-assembled monolayer/reduced graphene oxide- polyamidoamine dendrimer (SAM/ NH2rGO/ PAMAM) thin film was developed that detected the DENV-2 E-proteins with the detection threshold of 0.08 pM. This same research group also developed another sensor chip back in 2018 to detect the dengue virus \cite{omar2018development}. But in the later work, they introduced a graphene-oxide (GO) layer in the sensor chip that significantly enhanced the overall performance of the sensor. In a similar work, graphene-based material sensor chips were investigated for real-time and quantitative detection of DENV protein. In this study, the sensor chip was developed by accumulating cadmium sulfide quantum dots-reduced GO upon a thin gold plate. By changing the angle of incident light this graphene-based chip was able to detect DENV protein of concentration as low as 0.1 pM. Like dengue another deadly disease that causes hemorrhagic fever is ebola and certainly this virus has the potential to create another pandemic. Recently an SPR chip was developed to diagnose the ebola virus with high specificity and sensitivity \cite{sharma2020surface}. To develop the sensor a gold SPR chip was modified with 4-mercaptobenzoic acid (4-MBA). Three different monoclonal antibodies (mAb1, mAb2, and mAb3) of Ebola virus were in the race and the interactions of antibodies were investigated to determine the suitable mAb based on the affinity constant ($K_d$). After the screening mAb3 showed the highest affinity which later was confirmed by ELISA. This study also suggested the interaction was spontaneous, endothermic, and driven by entropy. Figure 3(a) shows a commonly used experimental setup generally known as Kretschmann configuration. Figure 3(b)-(d) illustrates the sensing schemes of different planar sensors.

\begin{figure} 

  \centering 

    \includegraphics[scale=.52]{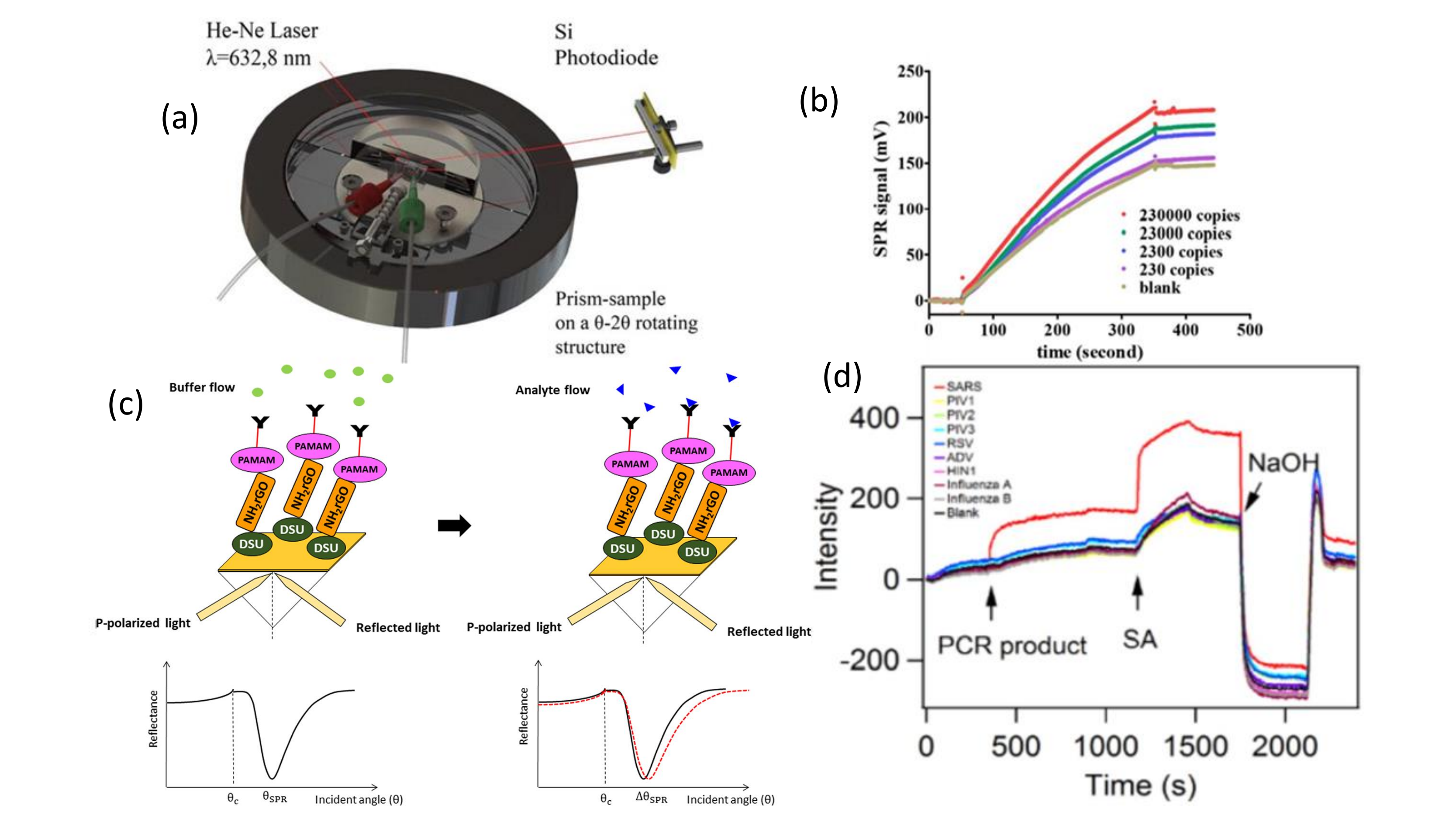} 

    \caption{(a) Experimental set-up in Kretschmann configuration for sensing (reproduced with permission from \cite{manera2012tio2}). (b) Signal amplitude of surface plasmon resonance varies in accordance with number viral bodies (reproduced with permission from \cite{chang2018simple}). (c) Antigen-antibody binding changes resonance condition (reproduced with permission from \cite{omar2020sensitive}). (d) Viral pathogen determined by SPR signal intensity with high specificity (reproduced with permission from \cite{shi2015development}).}
\end{figure}

In 2013 a new type of avian influenza H7N9 virus emerged in China, causing human infection with high mortality taking 612 lives. A quantitative and real-time diagnosis was crucial for eradicating the outbreaks of that emerging disease. A straightforward strategy for rapidly and sensitively detecting the H7N9 virus was using an intensity-modulated surface plasmon resonance (IM-SPR) biosensor integrated with a newly generated monoclonal antibody \cite{chang2018simple}. In another study, an SPR-based biosensor was developed for specific detection of nine common respiratory viruses including influenza A and influenza B, H1N1, respiratory syncytial virus (RSV), parainfluenza virus 1-3 (PIV1, 2, 3), adenovirus, and severe acute respiratory syndrome coronavirus (SARS) \cite{shi2015development}. A significant challenge in this work was amplifying viral bodies by PCR (Polymerase chain reaction). But an advantage was the same sensor chip could be used to diagnose multiple times after washing with NaOH solution.

\begin{figure} 

        \includegraphics[trim={1.5cm 0cm 0cm 0cm},clip,scale=.55]{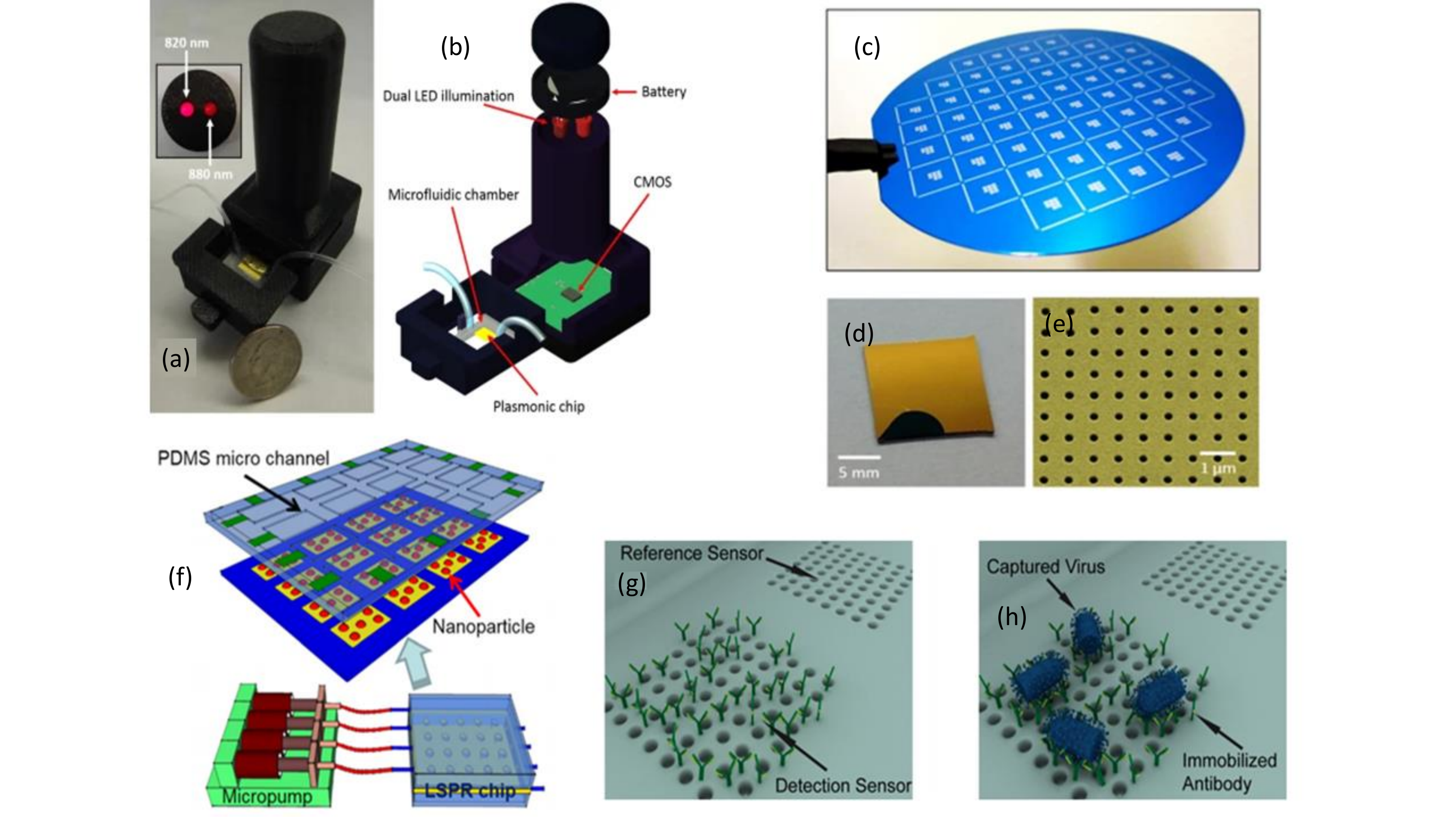} 

    \caption{(a) Portable small optofluidic biosensor. (b) Schematic diagram of portable biosensor using different wavelength LED.  (c) Photography of the wafer after deep-UV and dry etching steps. (d) Single plasmonic chip containing eight plasmonic pixels. (e) SEM image of the nanohole array with a hole diameter of 200 nm and an array period of 600 nm ((a)-(e) reproduced with permission from \cite{coskun2014lensfree}).(f) The illustration of LSPR sensor chip integrated with programmable microfluidics (reproduced with permission from \cite{geng2014route}). (g) Nano apertures holding antibodies. (h) Fluidic viral bodies being captured by immobilized antibody ((g)-(h) reproduced with permission from\cite{yanik2010optofluidic}).}   

\end{figure}

In a recent study, HIV virus was successfully detected by a commercially available simple planar SPR biosensor. HIV-related DNA with hairpin type DNA aptamers was diagnosed \cite{diao2018highly}. The proposed SPR biosensor could detect target DNA sensitively in a linear range from 1 pM to 150 nM with a detection limit of 48 fM. Lately, a typical planar SPR biosensor for medical diagnostics of human hepatitis B virus (hHBV) has been developed \cite{tam2017wide}. A 7-fold higher limit of detection and 2-fold increase in the coefficient of variance (CV) of the replicated results, were shown as compared to typical enzyme-linked immunosorbent assay (ELISA) testing.

\subsection{Optofluidic systems (plasmonic aperture systems)}
Another commonly used compact portable plasmonic biosensing platform is optofluidic systems. It is the combination of optoelectronics, optics, and nanophotonic with fluidics. Such a scheme constitutes a new perspective for manipulation of optical properties which incessantly scales the wavelength of light with applications ranging from fluidically adaptable optics to high sensitivity bio-detection. Many optofluidic plasmonic biosensors are based on nano apertures to enhance the plasmonic sensing capabilities of fluidic viral analytes. These nano apertures hold capturing molecules (e.g. antibody, aptamers) which increases binding potential between viral antigens and antibodies. One of the major recent breakthroughs in this regard is the detection of the COVID-19 virus\cite{funari2020detection}. In this device, the opto-microfluidic property is combined with plasmonic property. A serological testing with high specificity was devised. The refractive index (RI) sensitivity of the pure Au nanospikes in the opto-microfluidic device was precisely calculated by measuring the wavelength shift in the LSPR peak position when solutions with different refractive indexes (RI) are delivered to the microfluidic chip. The COVID 19 antibody presence was correlated with the LSPR wavelength peak shift of gold nanospikes caused by the local refractive index change due to the antigen–antibody binding. 

In 2010 nano-plasmonic biosensor chips were introduced which operated by utilizing the fluidic property of  the sample. Here the fluidic property was holding binding molecules by nano apertures. Detection and recognition of small enveloped RNA viruses (Vesicular stomatitis virus and pseudo typed Ebola) and large enveloped DNA viruses (vaccinia virus) was demonstrated. This platform opened opportunities for the detection of a broad range of pathogens in typical biology laboratory settings. 220nm radius aperture was formed in a metal-dielectric layer. These apertures hold capture molecules for selective viral detection\cite{yanik2010optofluidic}. In a more advanced work, programmable control systems for microfluidic flow of analytes were used to develop a more efficient and portable plasmonic biosensor. In this study, the 9 kinds of samples with different reflective index and antigen/antibody systems were utilized for characterization.  By using these programmable optofluidic arrays, the biomarker of liver cancer was tested \textit{in situ} and real-time.\cite{geng2014route}. 

Figure 4(a)-(e) exhibits a portable opto-fluidic sensor, which uses spectral shift of light to detect virus presence. Figure 4(f)-(h) illustrate how nano apertures hold immobilized antibody to capture pathogens.
  
\begin{figure} 

    \centering 

        \includegraphics[scale=.58]{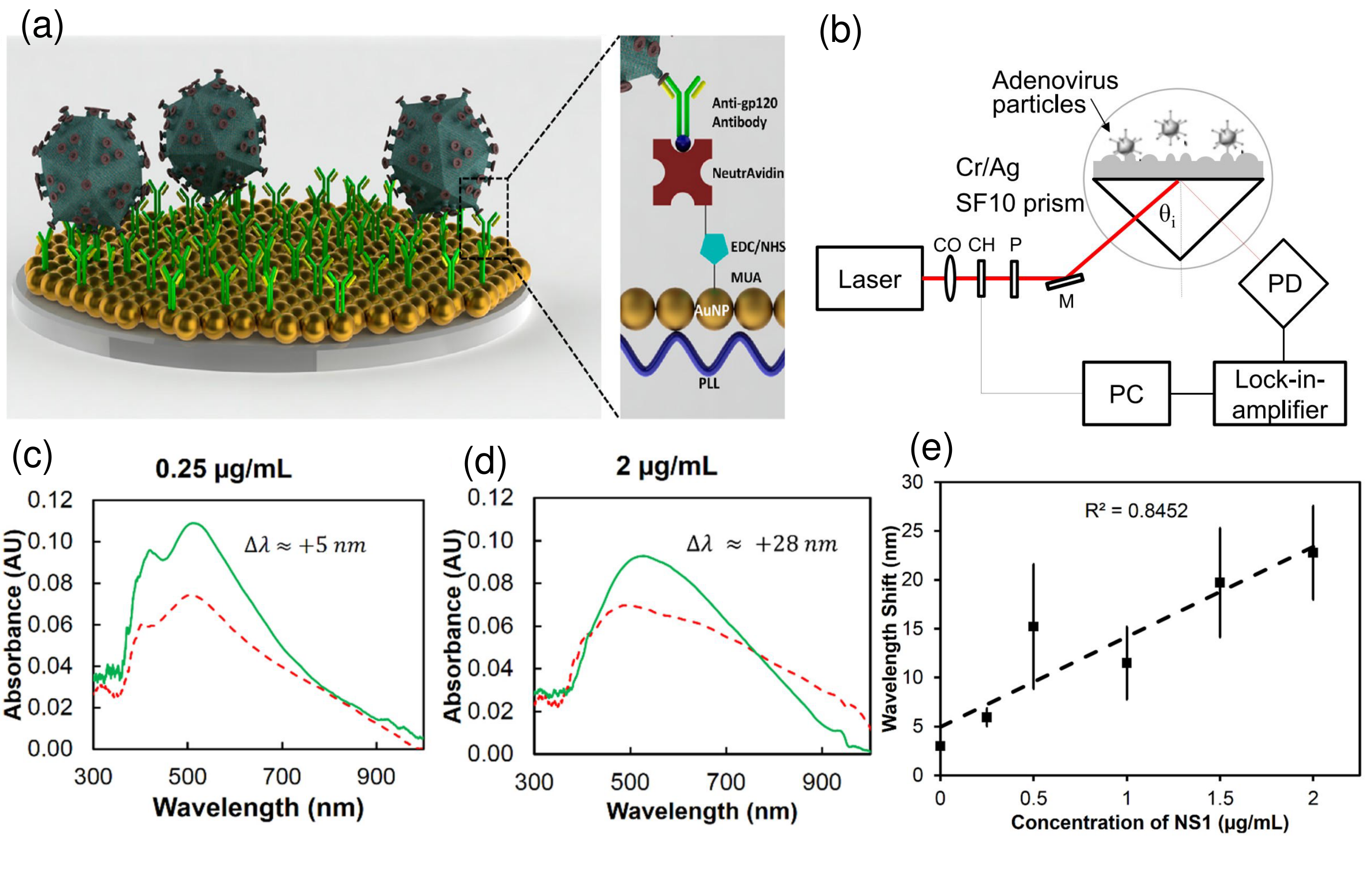} 

    \caption{(a) Nanoplasmonic biosensing platform for HIV detection (reproduced with permission from \cite{inci2013nanoplasmonic}). (b) Schematic diagram of optical set-up to detect adenovirus particles using gold nano island based plasmonic sensor (reproduced with permission from \cite{yu2013enhanced}). Absorbance spectra of the biosensor immobilized with anti-NS1 antibody (red dashed line) exhibiting wavelength shift of peak absorbance after incubation with plasma separated on-chip from whole blood samples containing (c) 0.25 $\mu$g/mL and (d) 2 $\mu$g/mL of dengue NS1 antigen (green solid line). (e) Plot of variation of wavelength shift with concentration of NS1 spiked into whole blood showing linearity with R2  value of 0.84 ((c)-(e) reproduced with permission from \cite{suthanthiraraj2019localized}).}

\end{figure}

\subsection{Nanoparticle enabled plasmonic structures}
Nanoparticles (NP) are primarily used in the plasmonic sensors due to their easy fabrication process and cost-effectiveness. Gold (Au) and silver (Ag) nanoparticles are used for this purpose. Au NPs’ SPR wavelength is found around 520nm and Ag NPs show LSPR usually around 400nm\cite{jans2012gold}. However, by changing the shape and size of the NPs, the SPR wavelength can be tuned\cite{lee2011small}. NP based plasmonic biosensors play a crucial role in virus sensing as they make label-free detection possible. As a result, no fluorescence or colorimetric biomarker is required. Different immobilizing antibodies are used to capture the viral particles or proteins. 

The first virus detection using nanoparticle based plasmonic sensor was reported by Fatih Inci et. al. in 2013\cite{inci2013nanoplasmonic}. They used Au nanoparticles with immobilized antibodies to detect and quantify different subtypes of HIV viruses from unprocessed whole blood. Their limit of detection was ${98 \pm 39}$ copies/mL for HIV subtype D. They measured the shift in resonant wavelength when the HIV virus was captured on the antibody immobilized biosensing surface as shown in Figure 5(a). They reported a maximum wavelength shift of 9.3nm for HIV sub-type A with virus concentration of 6.5 x $10^5$ copies/mL.  

During the same time, Yu et. al. detected adenovirus particles by using Ag nanoisland-based localized surface plasmon resonance\cite{yu2013enhanced}. It is to be noted that this was non-specific detection of adenovirus particles but based on numerical results they proposed specific detection models. They used rigorous coupled wave analysis and transfer matrix method using effective medium theory for numerical analysis. Change in reflectivity was measured to detect adenovirus particles as depicted in Figure 5(b) and the limit of detection was 109 viruses/mL. 

In 2013, Jahanshahi et.al. detected four diverse types of dengue virus using Immunoglobulin (Ig-M) based diagnostic test\cite{jahanshahi2014rapid}. They used Au coating on glass substrate to excite surface plasmons. Antigens and amines in the form of NPs were used to immobilize four different types of dengue virus. Change in reflection angle of SPR was measured to detect the viruses. They achieved sensitivity of 83-89\% in detecting the proven dengue cases and specificity was 100\%. Their LOD was 10 antibody titers. In 2016, Valdez and his colleagues used LSPR shift to detect respiratory syncytial virus (RSV) using gold, silver and copper nanoparticles \cite{valdez2016facile}. They used anti-RSV polyclonal antibody to bind the virus with metallic NPs and they found copper NPs perform better in detecting RSV with LOD of 2.4 PFU.

A year later, Lee et. al.\cite{lee2017binary} used Au NP and magnetic nanoparticle (MNP) decorated graphene (GRP) based hybrid structure to detect norovirus-like particles (NoV-LP). Enhanced plasmonic and electrical properties exhibited by this hybrid structure were used to detect NoV-LP.  The surface of Au/MNP-GRPs was functionalized with norovirus antibody to detect NoV-LPs and it performed well with high sensitivity and specificity. The change in resistivity was measured to detect NoV-LP in a concentration range from 0.01 pg/mL to 1 ng/mL and LOD was found to be 1.16pg/mL. At the same time, Kim and his group used gold nanoparticles in hetero-assembled sandwich format to detect hepatitis B surface antigen (HBsAg) using anti-HBsAg antibody as the binding antibody \cite{kim2018heteroassembled}. These sensors were very specific to HBsAg and the LOD was 100 fg/mL. 

In 2019, Heo et. al.\cite{heo2019affinity} used gold nanoparticle based LSPR biosensor to detect noroviral protein and human norovirus. They used for the first time norovirus recognizing affinity peptides to bind noroviral proteins which are relatively inexpensive compared to the binding antibodies. From the change in absorption value, they detected the presence of the virus. Their detection limit for noroviral capsid protein was 0.1 ng/mL and the limit of detection for human norovirus was 9.9 copies/mL. Lee and his group utilized LSPR method using Au spike like NPs to detect avian influenza virus using a multi-functional DNA 3 way-junction \cite{lee2019label}. They employed hemagglutinin (HA) binding aptamer and thiol group to bind the virus with NPs and achieved a LOD of 1 pM in two different environment of PBS buffer and diluted chicken serum.

At the same time, Sen et. al. \cite{suthanthiraraj2019localized} used thermally annealed thin silver film deposited onto silicon substrate to detect NS1 antigen of dengue virus in whole blood. Refractive index sensitivity of the biosensor was $10^{-3}$. A polyethersulfone membrane filter was used at the inlet of the sensor to separate blood cells from plasma and anti NS1 antibody was used to ensure specific binding of the NS1 antigen. An increase in absorption was found for antigen binding as shown in Figure 5(c)-(d) and for the highest 50 $\mu$g/mL concentration 108nm redshift in peak absorption wavelength was found. The sensitivity of this LSPR sensor was found to be ~9nm/($\mu$g/mL) and limit of detection was .06 $\mu$g/mL as shown in Figure 5(e). Recently in 2020, Qiu et. al. used dual functional plasmonic biosensor combining LSPR with plasmonic photothermal (PPT) effect to detect SARS Coronavirus 2 (SARS-CoV-2)\cite{qiu2020dual}.  They used functionalized two dimensional gold nano islands for sequence specific viral nucleic acid detection. Local PPT heat generated from Au NIs is used to transduce the in-situ hybridization for highly sensitive SARS CoV-2 detection and they achieved a detection limit of 0.22pM. During the same time, Rippa et.al. used two dimensional gold nano structures based on octupolar geometry to detect low concentration of rotavirus in water \cite{rippa2020octupolar}. They used rotavirus capsid (2B4) antibody to capture rotavirus for specific binding and detected the virus through the change in LSPR extinction wavelength shift. They achieved maximum wavelength shift of 46nm for a virus concentration of $10^5$ PFU/mL and they achieved LOD of $126\pm{3}$ PFU/mL.

\subsection{Quantum dots in plasmonic structures}

\begin{figure} 

    \centering 

        \includegraphics[scale=0.56]{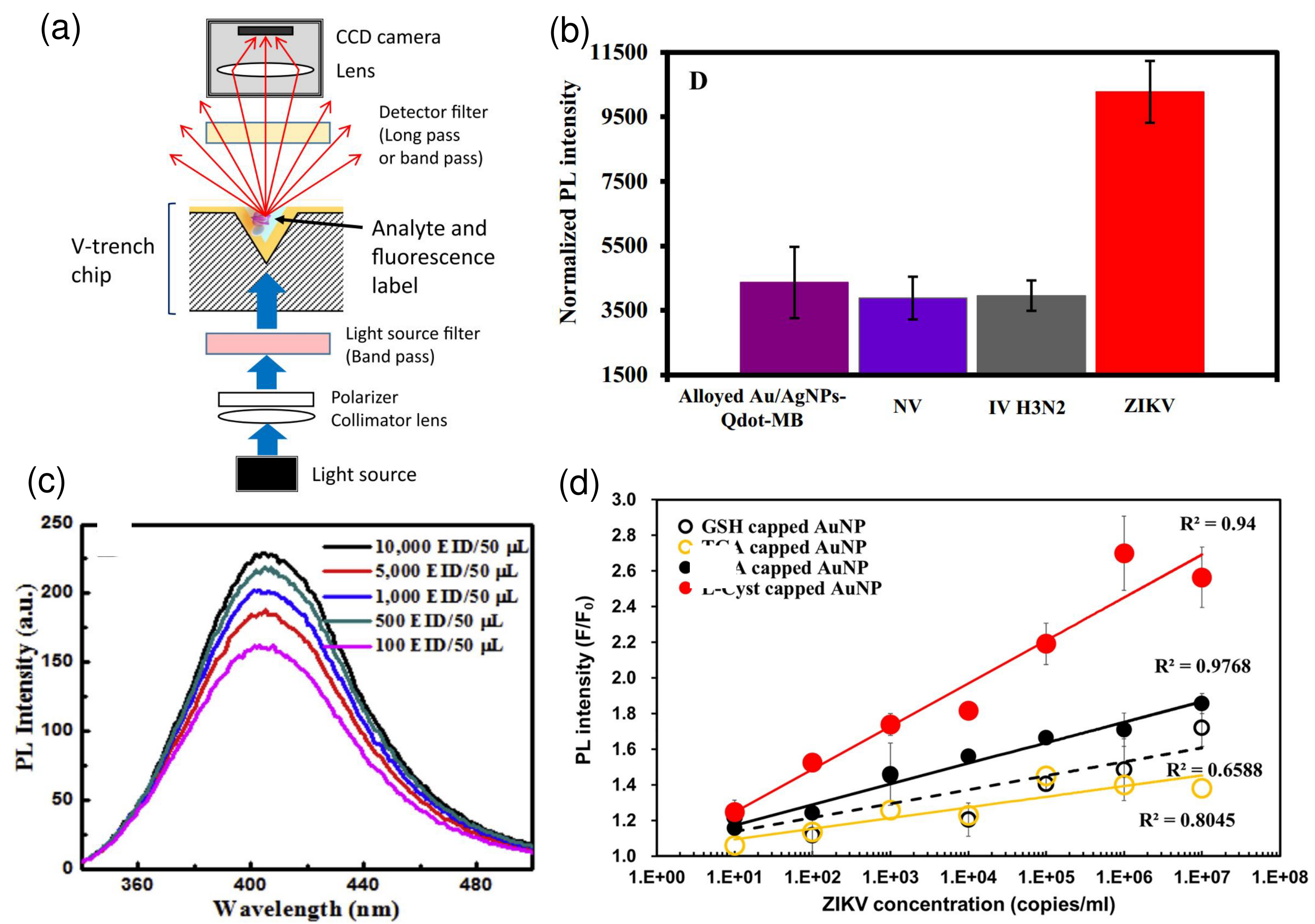} 

    \caption{(a) Schematic diagram of an optical system of a V-trench biosensor to detect norovirus like particles (reproduced with permission from \cite{ashiba2017detection}). (b) Selectivity of the plasmon alloyed AuAgNP-Qdot646-MB biosensor probe toward ZIKV RNA in the presence of the negative controls (influenza virus (IV) H3N2 and norovirus (NV)) (reproduced with permission from \cite{adegoke2017localized}). (c) Infectious bronchitis virus (Coronavirus) concentration dependent PL intensity of QD fluorescence biosensor (reproduced with permission from \cite{ahmed2018chiral}). (d) Calibration curves of ZIKV detection using 4 kinds of thiol-functionalized AuNPs. The LSPR signal amplifiers of L-cyst-AuNPs (red closed circles) and Ab-MPA-AuNPs (black closed circles) showed higher correlation coefficients than those of Ab-GSH-AuNPs (black open circles) and Ab-TGA-AuNPs (yellow open circles) (reproduced with permission from \cite{takemura2019localized}).}. 

\end{figure} 

Quantum dots(QDs) are used as fluorescence signal amplifiers to enhance the luminescent signals generated from the fluorescence probes attached to affinity reagents or viral targets. QDs generally have a core-shell structure and are made of inorganic substances like CdSe as core and ZnS as the shell. QDs possess exquisite optical properties and they exhibit stokes shift quantified up to hundreds of nanometers. Stokes shift is the difference between the maximum absorption wavelength and the maximum emission wavelength. QDs have several advantages like tunable broad absorption and narrow emission spectra and unlike traditional fluorescent dyes they can detect multiple signals simultaneously. Xuepu Li et. al.\cite{li2012fast} used quantum dot fluorescence along with gold nanoparticles to detect Avian Influenza Virus (AIV). They did not use plasmonic biosensor in this work but their work paved the way for using QDs as fluorescent signal enhancer in plasmonic biosensors virus detection.

In 2016, Takemura et.al. reported LSPR induced immunofluorescence nano biosensor by using CdSeTeS QDs with Au NPs to detect influenza virus \cite{takemura2017versatility}. Here quaternary CdSeTeS QDs were used to enhance the fluorescent signal generated from the antibody antigen interaction on the thiolated Au NPs. Anti-neuraminidase antibody and anti-hemagglutinin antibody were conjugated with thiolated Au NPs and quaternary QDs respectively. They achieved a limit of detection of 0.03 pg/mL for H1N1 influenza virus in deionized water, 0.4 pg/mL for influenza H1N1 virus in human serum and 10 PFU/mL for clinically isolated H3N2 virus. During the same time,  surface plasmon resonance assisted CdSe-ZnS based quantum dots were first used to detect norovirus like particles \cite{ashiba2017detection}. Excitation wavelength of 390nm was used to excite SPR on an Al film of the sensor chip equipped with a V-shaped trench as shown in Figure 6(a). To immobilize proteins a self-assembled monolayer (SAM) of phosphonic acid derivative was used. Concurrently gold and silver plasmonic nanoparticles with semiconductor quantum dots were used by Adegoke et. al \cite{adegoke2017localized} to detect Zika virus RNA. They used four different plasmonic NPs functionalized with 3-mercaptopropionic acid (MPA). MPA-AgNPs, MPA-AuNPs, core/shell (CS) Au/AgNPs, and alloyed AuAgNPs along with CdSeS alloyed Qdots were used to form the respective LSPR-mediated fluorescence nano biosensor the PL intensity of which is shown in Figure 6(b). They achieved minimum LOD for alloyed Au Ag NP which is 1.7 copies/mL and it was very selective toward Zika virus RNA. 

In 2018, Ahmed and his colleagues \cite{ahmed2018chiral} proposed and demonstrated a novel method by using Zr NPs with Zr QDs to detect Coronavirus. Zr QDs show blue fluorescence emission and by functionalizing them with anti-infectious bronchitis virus (IBV) antibodies, anti-IBV antibody-conjugated magneto-plasmonic nanoparticles (MPNPs) are formed. As shown in Figure 6(c), from the change in photoluminescence intensity they detected Coronavirus and their LOD was 79.15 EID/50$\mu$L. In another work, an immunofluorescence biosensor for the detection of nonstructural protein 1 (NS1) of the ZIKV by using gold NPs and QDs was made \cite{takemura2019localized}. The LSPR signal from the Au NPs was used to amplify the fluorescence signal intensity of quantum dots (QDs) from the antigen-antibody detection process. CdSeTeS QDs were used with four different thiol capped Au NPs and their PL intensity with different ZIKV concentration is shown in Figure 6(d). The biosensor achieved a LOD of 8.2 copies/mL and could detect the virus within the concentration range $10-10^7$ RNA copies/mL. It could detect the ZIKV in human serum and showed good specificity for NS1 antigen against other negative control targets. At the same time, Omar et. al used CdS QDs with amine functionalized graphene oxide thin film to detect the dengue virus E-protein \cite{omar2019sensitive}. They used monoclonal antibodies (IgM) to bind the protein and achieved an outstanding LOD of 1 pM.  

Very recently, Nasrin and her group \cite{nasrin2020fluorometric} employed CdZnSeS/ZnSeS QD-peptide and gold nanoparticle composites to enhance the LSPR signal to detect different concentrations of influenza virus from $10^{-14}$ to $10^{-9}$ g/mL. They varied fluorescence intensity by changing the distance between the QD and NP by using different peptide chain lengths to find the optimized condition to detect the virus and achieve a detection limit of 17.02 fg/mL. Previously, the same group detected norovirus \cite{nasrin2018single} using the same mechanism and their LOD was 95 copies/mL. However, this system could not detect the small change in norovirus concentration due to the smaller crosslinker between two NPs.

\subsection{Nanowire and nanorod based plasmonic biosensor}
\begin{figure}
	\centering
		\includegraphics[trim={3.5cm 1cm 4cm 0.8cm},clip, scale=0.8]{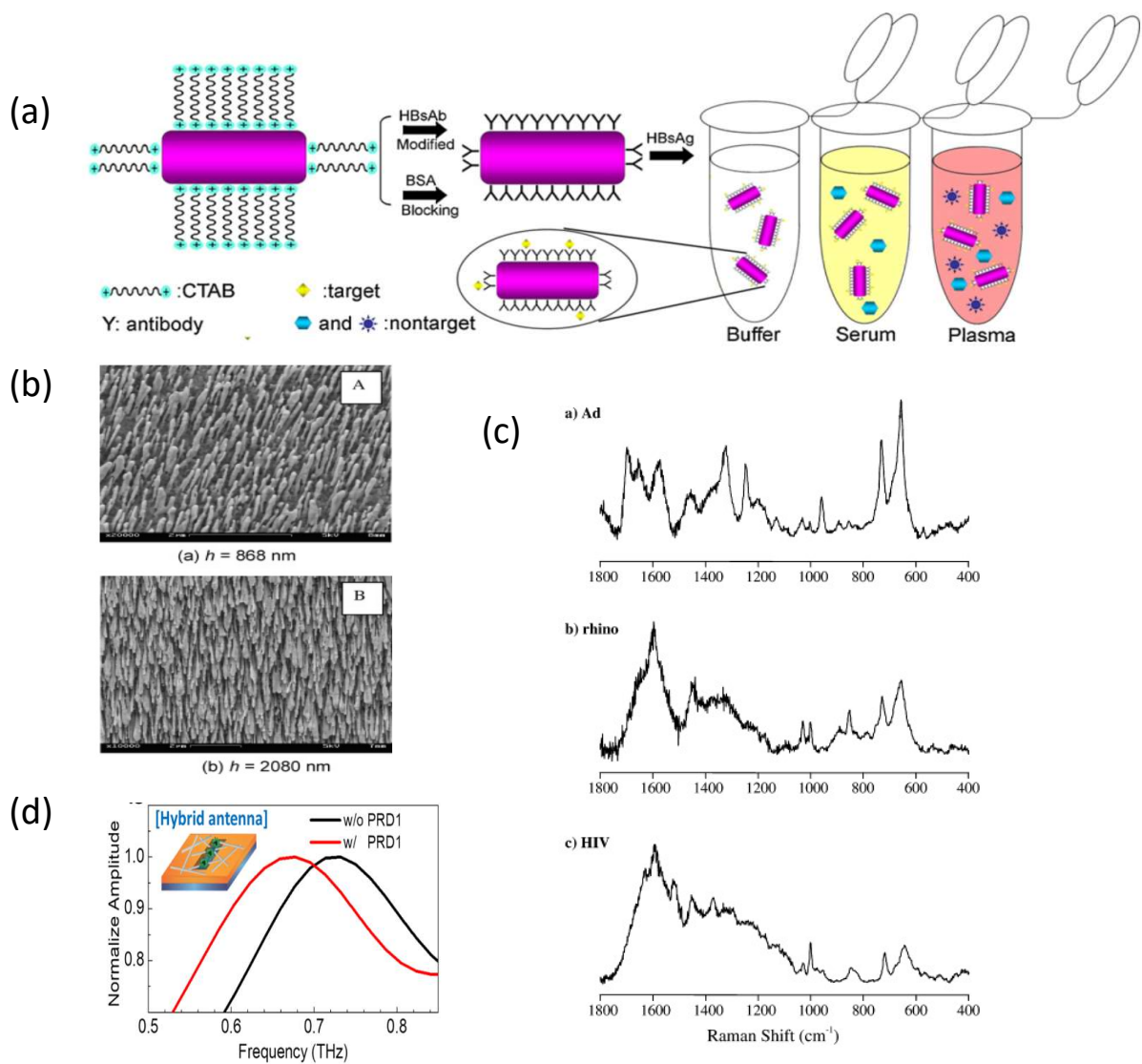}
	\caption{(a) Schematic representation of the synthesis of HBsAb antibody functionalized gold nanorods and the detection mechanism for the biosensor immunoassay in capturing targets in different matrices (reproduced with permission from \cite{wang2010gold}). (b) Representative scanning electron micrographs of the Ag nanorod arrays deposited with different lengths, (A) h = 868 nm and (B) h = 2080 nm. The typical SERS substrate used for virus detection is represented in (A) i.e. 870 nm (reproduced with permission from \cite{shanmukh2006rapid}). (c) Representative SERS spectra of RSV, HIV, adenovirus and rhinovirus (reproduced with permission from \cite{shanmukh2006rapid}). (d) Normalized transmission amplitudes through hybrid slot antenna (lNW= 1$\mu$m) with (red) and without (black) PRD1 viruses (reproduced with permission from \cite{hong2018enhanced}).}
	\label{NW}
\end{figure}

Nanowires and nanorods are usually used to enhance the properties of existing biosensors due to their ability to confine electromagnetic fields in a superior way.

Das and his colleagues designed and simulated a plasmonic immunoassay, in 2020, which comprised of sandwich plasmonic biosensor whose sensitivity was enhanced by using gold nanorods. They have varied the prismatic configuration and found that the BK-7 glass-based sensor has a sensitivity of 111.11 deg/RIU \cite{das2020gold}. They also varied the distance of the gold NRs and their aspect ratio and recorded their observations. This sensor was designed for SARS-COV-2 detection, the gold nanorods (NRs) and gold nanosheets are functionalized with SARS-COV-2 spike-protein antibodies, and shift is observed in the SPR angle.
In 2010, a biosensor was devised by functionalizing gold nanorods with monoclonal hepatitis B surface antibody (HBsAb) through physical adsorption. The characteristic plasmonic absorption spectrum is then measured after placing these nanorods in the vial containing blood serum. The LOD of the sensor is 0.01 IU/ml \cite{wang2010gold}; the schematic representation is shown in Figure \ref{NW} (a).

In 2006, a biosensor was fabricated by Shanmukh et al. where they obliquely deposited silver nanorods on the surface, the scanning electron micrograph is shown in Figure \ref{NW} (b). Three different human viruses were detected by the device namely adenovirus, rhinovirus, and HIV. After depositing the virus, Raman spectrum is measured and from the change in the spectrum, viruses can be detected. This sensor mainly results in improved SERS detection; Ag nanorod
substrates exhibit extremely high ($\sim$10$^8$) SERS enhancement factors. The SERS spectra of different viruses are shown in Figure \ref{NW} (c). The vibrational spectra of the molecule adsorbed on the sensor surface are enhanced as the incoming laser beam interacts with the electrons in the plasmonic oscillation in the nanorods \cite{shanmukh2006rapid}. 
In 2018, Hong et al. developed hybrid slot antenna structures in the THz frequency range, where silver nanowires (AgNWs) were employed to increase the sensitivity. The schematic representation of this structure is shown in Figure \ref{NW} (d). They used this structure for virus detection. PRD1 bacteriophage virus was detected using this sensor. The THz spectrum before and after placing the virus was used for the detection purpose \cite{hong2018enhanced}.

\section{Metamaterial based biosensors}
\subsection{Sensing principles of metamaterial based plasmonic biosensors}

Metamaterials are engineered materials; they possess properties that are not found in naturally occurring elements \cite{kshetrimayum2004brief}. These exotic properties depend on the geometry of the material hence can be tuned as per requirement. Usually, specific repeating patterns are created on the material and each pattern has size smaller than the wavelength they need to affect.
 
Metamaterials enable the detection of biomolecules in THz-GHz frequency regimes which is difficult otherwise, as microorganisms such as fungi, bacteria, and viruses have scattering cross-sections that are much smaller than THz/GHz wavelengths. Sensing biomolecules in the THz-GHz electromagnetic spectra has several advantages as it provides label-free, non-contact, and non-destructive sensing.
 
\begin{figure} 

    \centering 

        \includegraphics[scale=1]{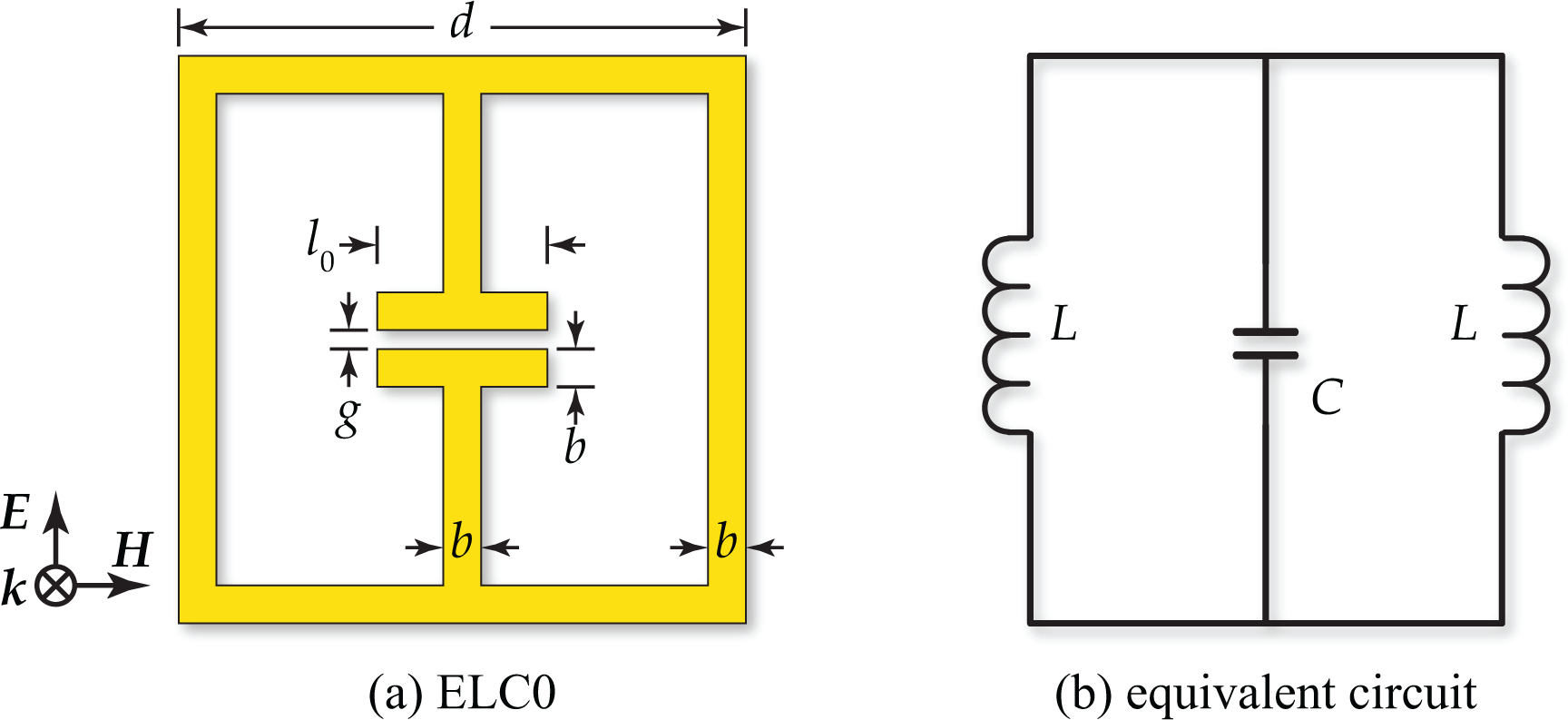} 

    \caption{LC equivalent circuit for a metamaterial (reproduced with permission from \cite{Withayachumnankul:10}). } 
    \label{equimeta}

\end{figure}

Metamaterial was first used in biosensing by Lee et al. In 2008 \cite{lee2008biosensing}. They used a gold split ring resonator (SRR) array to detect biotin and streptavidin to show the biosensing capability of the metamaterial. SRR worked as a biosensor as it can be considered as a simple LC circuit with a simple resonant frequency of
\begin{equation}\centering
    f_0 \approx \frac{1}{2\pi\sqrt{LC}}
    \label{meta}
\end{equation}
So, the resonant frequency of the system changes as the capacitance or inductance changes. As biotin and streptavidin bind to the system the capacitance of the SRR changes which is reflected in the resonant frequency thus, it can be used as a biosensor.
In Figure \ref{equimeta} the equivalent LC representation of a specific metamaterial with rectangular geometry with a split is shown. In this way, the metamaterials with nanogap can be modeled and their sensing mechanism can be understood using equation \ref{meta}.


\subsection{Metamaterial based biosensors for virus detection}

\begin{figure}
	\centering
		\includegraphics[trim={0cm 1cm 0cm 1cm},clip,scale=0.58]{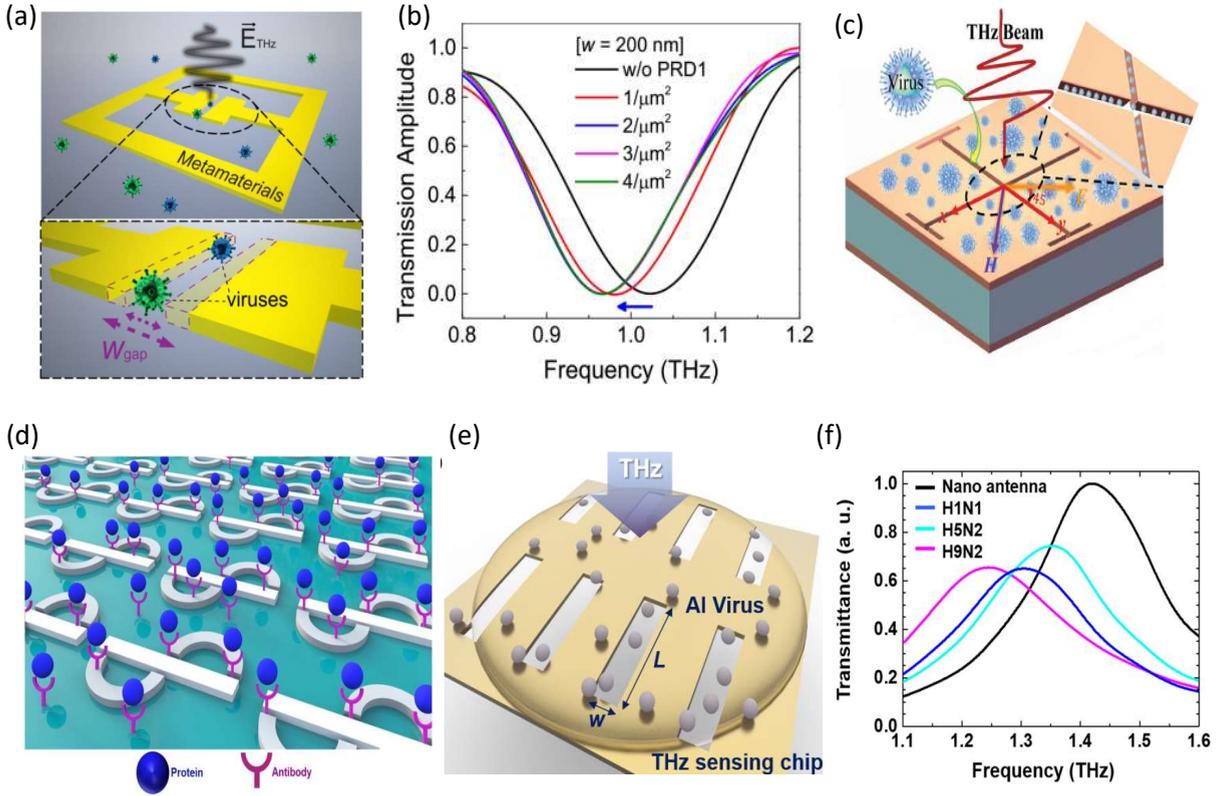}
	\caption{(a) Schematic of THz nano-gap Au metamaterial sensing of viruses (reproduced with permission from \cite{park2017sensing}). (b) Normalized THz transmission amplitudes of THz metamaterials after deposition of PRD1 at various surface densities for gap width of  200 nm  (reproduced with permission from \cite{park2017sensing}). (c) 3D schematic diagram of the THz biosensing  metamaterial absorber for AI virus detection (reproduced with permission from \cite{cheng2018terahertz}). (d) Schematic demonstration of ZIKV envelope protein binding with respective antibody on the toroidal THz plasmonic metasurface (reproduced with permission from \cite{ahmadivand2017rapid}). (e) Schematic of THz detection of virus samples in liquid state using a nano slot-antenna array based sensing chip (reproduced with permission from \cite{lee2017nano}). (f) Transmittance spectra through multi-resonance nano-antenna  with and without virus sample onto the antenna (reproduced with permission from \cite{lee2017nano}).}
	\label{mmimage}
\end{figure}

In 2017, S.J. Park and his group created a metamaterial surface using a gold rectangular structure on quartz, the schematic of the structure is shown in Figure \ref{mmimage} (a). Metamaterial has certain capacitance and inductance equivalent parameters. The presence of virus particles within the capacitor gap changes the resonance frequency which can be explained by a simple LC circuit. Hence different viruses were detected by observing the THz transmission spectra. They detected bacteriophage viruses PRD1 (60 nm) and MS2 (30 nm). Sensitivity was 80 GHz/ particle \cite{park2017sensing}; the transmission spectra of the biosensor after deposition of virus at various surface density is shown in Figure \ref{mmimage} (b).
Ahmadivand and his colleagues designed a toroidal metamaterial-based biosensor that detected zika virus envelope protein by measuring the spectral shifts of the toroidal resonance in 2018. They also added gold nanoparticles to see the effect in the sensitivity and observed enhancement in the performance of the sensitivity \cite{ahmadivand2018extreme}.
In 2018, A THz biosensing metamaterial absorber for virus detection based on Spoof Surface Plasmon Polariton (SSPP) Jerusalem cross apertures metamaterial absorber was devised. They determined the shift in absorption and resonant frequency as the alpha-beta parameters of the viruses were changed. Analyte thickness of H9N2 was also changed to see the variation in resonant frequency and absorption. From this observation, they claimed that virus subtypes can be uniquely identified using this sensor. H5N2, H1N1, H9N2 viruses were detected \cite{cheng2018terahertz}. The schematic is shown in Figure \ref{mmimage} (c).

In 2017, Ahmadivand and his group used 2D microstructures composed of iron (Fe) and titanium (Ti) for the magnetic and electric resonators (torus), respectively to design a set of asymmetric split resonators as meta-atoms to support ultra-strong and narrow magnetic toroidal moments in the THz spectrum; the schematic demonstration of this structure is shown in Figure \ref{mmimage} (d). Limit of detection of ~24.2 pg/mL and sensitivity of 6.47 GHz/log(pg/mL) their system resulted in toroidal response lineshape extremely sharp, narrow, and deep. They analyzed the sensitivity of the dip with Zika virus envelope protein attached to the system \cite{ahmadivand2017rapid}.
In 2017, Lee et. al. used a multi-resonance and single resonance nanoantenna sensing chip which was fabricated using gold nanoantennas printed on a silicon wafer to sense different types of Avian Influenza viruses. H9N2 was sensed using a multi resonance sensor. By using a single resonance nanoantenna they demonstrated that viruses can be classified in terms of resonance frequency and decreased transmission ratio \cite{lee2017nano}. The schematic of the biosensor is depicted in Figure \ref{mmimage} (e) and the transmission spectra with and without virus particles are shown in Figure \ref{mmimage} (f).

In 2019 Vafapour and his colleagues developed a biosensor using metamaterial comprising of H-shaped graphene resonator on a semiconductor film which they used to detect Avian Influenza \cite{8663385}.
Ahmed et al. developed a cost-effective metasurface-based biosensor in 2020. They used a Digital Versatile Disc which already has built-in periodic grating where they deposited multilayers of gold, silver, and titanium and showed that the device exhibits fano-resonance. When the HIV particles were captured they observed a shift in the fano resonance peak from which HIV can be detected \cite{ahmed2020tunable}

In 2016, Aristov et al. devised a 3D metamaterial-based biosensor composed of woodpile structure which has not yet been used in virus detection but showed sensitivity greater than 2600nm/RIU and phase-sensitive response is more than $3\times10^4$ degrees/RIU for analytes which is very high. In the same year, Sreekanth et al. designed a biosensor with grating coupled hyperbolic metamaterial which is a bulk 3D sub-wavelength structure that enhances the angular sensitivity of the plasmonic biosensor. They detected cowpea virus with it and obtained a maximum sensitivity of 7000$\deg$ per RIU \cite{sreekanth2016enhancing}.

\includepdf[pages={1}]{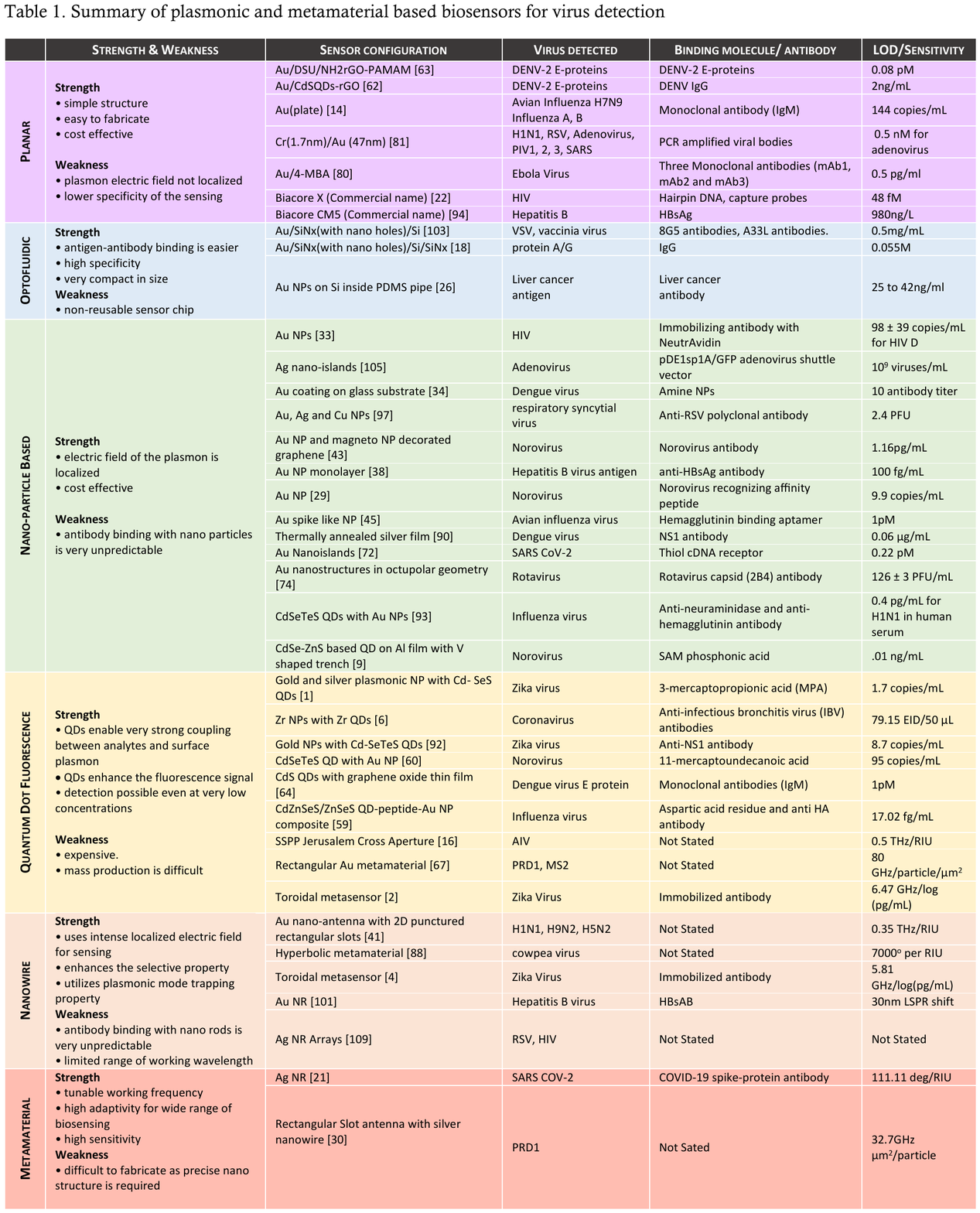}

\section{Plasmonic Biosensors for Coronavirus Detection}

Currently, diagnosis of COVID-19 is primarily accomplished by three techniques- quantitative reverse-transcription polymerase chain reaction (RT-qPCR) \cite{corman2020detection} and gene sequencing, a lateral flow immunoassay, which is a common point-of-care (POC) diagnostic approach that detects antibodies against SARS-CoV-2 in patient samples \cite{boger2020systematic,bastos2020diagnostic}, and  chest computed tomography (CT) \cite{zhang2020clinically}. Quantitative reverse transcription polymerase chain reaction (RT-qPCR) is widely used as the confirmatory test for COVID-19 detection and it is considered as the gold standard in this regard. However, the RT-qPCR method is lengthy and involves difficult processing methods. It demands highly trained manpower and cost which hinder the large-scale testing for COVID- 19. Although the RT-qPCR test is highly sensitive, may give false negative reports especially if the specimen is collected from the upper respiratory tract after a certain period from the onset of symptoms \cite{tahamtan2020real}. Therefore, there is an ongoing demand for an alternative detection method for novel coronavirus.

Plasmonic biosensing is a promising field for the detection of Coronavirus, as it can enable rapid testing and also reduce the manpower needed for performing the diagnosis. There are already many ongoing and reported works on plasmonic detection of novel Coronavirus. The schemes that were developed for detecting other coronaviruses can be useful for SARS-CoV-2 detection as well. Commercially available surface plasmon resonance (SPR) and localized surface plasmon resonance (LSPR) sensors are already being used for viral strain detection such as SARS, MERS, and influenza \cite{bhalla2020opportunities}.
In May 2020, Moitra et. al. reported work on COVID-19 detection using plasmonic nanoparticles which produced results within 10 minutes \cite{moitra2020selective}. Ahmadivand et. al. reported in June 2020 their work using toroidal plasmonic metasensor for femto-molar detection of the COVID-19 spike protein. They claimed that their sensor had a limit of detection of around 4.2 fmol and sample-to-result duration was around 80 minutes \cite{ahmadivand2020femtomolar}. In the same year, Das et al designed a gold nanorod-based plasmonic sensor to detect COVID-19 with a sensitivity of 111.11 deg/ RIU \cite{das2020gold}.  In 2018, Ahmed et. al. devised a sensor using magneto-plasmonic nanoparticles for coronavirus detection. The fluorescence properties of immuno-conjugated QD MP NPs nanohybrids through separation by an external magnetic field enabled biosensing of Coronavirus with a limit of detection of 79.15 EID/50 mL \cite{ahmed2018chiral}.  In August 2020 Uddin et. al. proposed a surface plasmon resonance sensor for COVID-19 detection which they claimed to have a sensitivity of 130.3 degrees/RIU \cite{uddin2020numerical}. Previously, in 2009, Huang et. al developed a localized surface plasmon coupled fluorescence fiber-optic biosensor for the detection of SARS-CoV \cite{huang2009detection}. Funari et al. devised a detection system that combined opto-microfluidic chip with LSPR to detect antibodies against SARS-COV-2 spike protein in December 2020 \cite{funari2020detection}.
In addition to that, several works have been mentioned in this article in which viruses with a diameter smaller than coronavirus (120 nm) are detected e.g. PRD1 (60 nm) MS2 (30 nm), Avian Influenza Virus (80 nm - 120 nm), Zika Virus (50 nm), etc. using plasmonic and metamaterial-based biosensors with promising sensitivity, these schemes might be useful for the detection of Coronavirus as well and further investigation can be done to prove their usefulness in coronavirus detection. It is noteworthy that there is still no reported work of Coronavirus detection using metamaterial-based biosensors.

\section{Future perspective}
Researchers are always on the quest of finding new methods of pathogen detection which is faster, accurate, sensitive, and also can be used as a point-of-care device because the population is increasing at a fast pace and epidemics are materializing more frequently than ever. In recent times, carbon-based materials like graphene \cite{cucci2019hybrid,gong2019hybridization}, carbon nanotube\cite{wang2020ultrahigh}, and nanodiamond \cite{miller2020spin} are introduced in biosensing. 

Like other fields, machine learning has garnered significant interest from researchers in the field of biosensing. Recently the behavior of plasmonic biosensor designs that integrate metamaterials based on machine learning algorithms was explored \cite{moon2020machine}. The overall design process was defined by two steps. The first step involves the initialization of classical calculation before machine learning. The performance of various detection scenarios was tested with sensing of different DNA oligomers. The results were employed to generate both training and test sets. In the second step, a machine learning algorithm was applied to predict the performance of metamaterial-based plasmonic biosensors. The use of meta-plasmonic structure with the help of machine learning has enhanced the detection sensitivity by more than 13 times. Lately, another work has also shown the use of diverse machine learning algorithms that aims to improve the quality of the real-time SPR responses called sensorgrams \cite{gomes2021smartspr}. In a different work, deep learning methods enabled by artificial neural networks were used as a powerful and efficient tool to construct correlation between plasmonic geometric parameters and resonance spectra\cite{li2019deep}. With this numerical method,  the spectra of millions of different nano-structured sensors can be predicted. 

Topological insulators have emerged as new building blocks in photonics and they offer promising prospects for plasmonic biosensing. It is one kind of new material and bulk insulator with a chiral Dirac cone on its surface. It is a new state of quantum matter with internal insulation but the electric conduction for the interface. Recently these are being used to enhance the sensitivity and detection limit of surface plasmon resonance based sensors \cite{zhu2019topological,zhao2019high}. 

Quantum enhanced plasmonic sensing is a state of the art idea as quantum properties of light adds more degrees of freedom in biosensing. Quantum properties can amplify the sensitivity of a sensor and thus has the potential to significantly improve the plasmonic sensing scheme through the development of quantum-enhanced sensors. For this purpose, recently researchers fabricated an array of sub-wavelength nano-structured holes in a thin silver film. This sensor used an effect known as extraordinary optical transmission (EOT) \cite{dowran2018quantum}. This phenomena preserved the quantum properties of the light and made the use of quantum states of light, a viable option to enhance the sensitivity of plasmonic sensors. 

 Surface plasmon resonance imaging (SPRi) is another emerging field that has been applied in the detection and monitoring of biomolecular events \cite{puiu2016spr}. There has been a study of the apple stem pitting virus (ASPV) by imaging the aptamer binding with the coat protein using SPR \cite{lautner2010aptamer}. Recently, another SPRi based technique, named ultra-near-field index modulated PlAsmonic NanO-apeRture lAbel-free iMAging (PANORAMA) was developed\cite{ohannesian2020plasmonic}. It relies on unscattered light to detect dielectric nanoparticles. PANORAMA can detect, count and size individual nanoparticles beyond 25 nm, and dynamically monitor their distance to the plasmonic surface at a millisecond timescale. 

Finally, the sensitivity of plasmonic and metamaterial-based biosensors has reached femtomolar detection limit but here limiting factor is the specificity of the biosensors which requires attention. Although, by employing aptamer and peptide-based binding molecules specificity has improved significantly, their application is still not suitable for all biosensors. Moreover, the biosensors may detect viruses successfully in a laboratory environment, their performance needs to be evaluated from clinical samples. Another issue for metamaterial-based biosensors can be mass production for which the sensing platform needs to be scalable for reduced fabrication cost. Plasmonic biosensors can pave the way for achieving multiplexing capability for detecting multiple viruses \cite{sanchez2017surface} using the same sensing platform. 

\section{Conclusion}
The world is currently fighting the pandemic caused by SARS-CoV-2, there is no assurance when will this pandemic end let alone the next one. Disease diagnosis in the early stage is one of the main weapons in this ongoing fight against the pandemic. Though during the last few years there has been significant improvement in disease diagnosis by optical biosensors, even COVID-19 has been successfully detected by LSPR based biosensors there is still room for development. Plasmonic and metamaterial-based biosensors exist in many different forms and each form has one or more supremacy over conventional techniques, some are already being used in laboratories for drug and vaccine developments \cite{myszka2000implementing}. Some of the plasmonic biosensors like planar metal-dielectric interface-based biosensors have simple fabrication techniques and give pretty good sensitivity and low LOD. Plasmonic biosensors like those based on metamaterials and nanowres allow label-free, non-destructive sensing. Nanoparticle based plasmonic biosensors allow a broad range of antibody binding and metamaterial based biosensors have increased the sensitivity manifolds. Most importantly all the biosensors make rapid detection possible. However, plasmonic and metamaterial-based biosensors need to be robust and reproducible to become mainstream virus-caused disease diagnosis methods. Also, biosensors need to be developed as a lab-on-a-chip system to make them ubiquitous. Many of the researchers are already working on a lab-on-a-chip configuration of plasmonic biosensors. If these shortcomings can be overcome in near future then plasmonic and metamaterial-based biosensors will enable faster and more accurate detection of pathogens which will greatly help to prevent outbreaks in the future.



\printcredits

\bibliographystyle{cas-model2-names}

\bibliography{main}


\end{document}